# Understanding Time and Causality is the Key to Understanding Quantum Mechanics

William R. Wharton
Physics Department, Wheaton College, Wheaton, Illinois 60187

**Abstract**
A new interpretation of quantum mechanics, similar to the Copenhagen interpretation, is developed from time-symmetry arguments and commonly held principles concerning time and causality. These principles, which are grounded in ideas outside of quantum mechanics, suggest that the strange features and paradoxes of quantum mechanics come from backward causation, in which future events can change the past. Using these principles this paper gives a better understanding of (and reasons for) stationary states, tunneling, wave-particle duality, the measurement problem with the collapse of the state vector, Hardy's paradox, non-locality, and the EPR paradox. These are only a few of the interpretational successes, and the model will be contrasted to other popular and/or recent interpretations. Unfortunately, this model is metaphysical, without any predictive power.

## I.     Introduction

The success of the formalism of quantum mechanics, QM, in describing and predicting properties of nature is unmatched by any other theory. Nevertheless every interpretation of QM is fraught with incompleteness, complexities, ugliness, and/or contradictions. The complexities often take the form of imagining new undetectable/ unknowable components of nature, such as pilot waves, hidden variables, or multi-universes. A more elegant approach is to take what we already know about nature and show how it can naturally account for the strange features of QM. Ugliness manifests itself in symmetry violating processes, particularly time-symmetry violation, such as in the sudden collapse of the wavefunction. A more beautiful solution keeps the symmetries. The incompleteness problem is a failure to explain or account for all of the strange paradoxes of QM. This failure is accepted reticently by a positivist or instrumental philosophy, with the limited goal of empirical adequacy, known as FAPP, "for all practical purposes". Questions about QM, which are asked in ignorance are discarded with the statement[1], "Ask a foolish question and you will get a foolish answer." The contradiction problem refers to an interpretation of QM violating well-founded principles existing in physics outside of QM. The two principles, which are most commonly trampled upon are locality, as it exists in special relativity, and the electro-magnetic theory prediction that an accelerating electric charge must be emitting or absorbing electro-magnetic radiation. For example, stationary states of atoms do not continuously emit or absorb radiation. Many QM interpretations of these states imply that the orbiting electrons have some sort of momentum, even if it isn't well defined, and that they are continuously accelerating because of the Coulomb force. The purpose of this paper is to show that the clarification of the meaning of time and causality, using widely accepted ideas existing outside of QM, will go a long way toward alleviating all of these problems associated with understanding QM.

Many consider the measurement problem, defining what constitutes a measurement of a quantum system, as the most pressing problem in the interpretation of QM. Quantum systems interact with their environment, but only a select few of these interactions constitute a measurement. A measurement, in contrast to all other interactions, selects a particular value, out of multiple possible values, for an observable of the quantum system. The majority of interpretations claim that before the measurement, the observable didn't have a specific value. All non-measurement interactions, which are dependent upon this observable, require a superposition of multiple values for the observable. This multi-valued result is known as entanglement between the two objects doing the interaction. The underlying reasons for the difference between measurements and other types of interactions, involving entanglement between objects, are very easily and naturally explained using the principles of time and causality. These widely accepted principles are presented below and only minimally justified within the manuscript. Using these principles, we will develop a less complex, more beautiful, and more complete interpretation of QM, which is superior, based on standards elucidated above,to all previous interpretations. The principle of causality, given as basic principle #2, will identify a likely source of incompleteness within QM. Whereas most interpretations of QM choose to relax the universality of the principle of causality, this paper will instead hold fast to the principle of causality and relax the universality of QM.

**II.    Basic Principles**
1. An event, b, is a point like reality at a space-time location $x_b$, $y_b$, $z_b$, $t_b$.
2. Any event must either be a "first cause" or embedded in a causal chain, in which cause-effect is propagating in a unique direction through a series of events. The former case initiates a new causal chain and the latter case means the event in question is influenced by some earlier event(s), and in-turn influences different events further down the causal chain.
3. Time is a coordinate, like the spatial coordinates, and doesn't have any flow. Specifically, different coordinate systems have different times for the same event, and the transformation of time from one frame to another depends upon the spatial coordinates. Concepts of past, present, and future are not an inherent property of time, since time doesn't have any flow.
4. The flow, which we associate with the concepts of past, present, and future is a causal flow, which doesn't depend upon the reference frame, i.e. each causal chain flows the same direction in all reference frames. However, different observers have different definitions of the present, because they are experiencing different causal chains. An observer, or agent, has present knowledge of his/her past, which is the collection of all events earlier in his/her causal chains. The agent is acting in his/her present to have a causal effect on his/her future, which is further down the causal chain(s). Since different observers are experiencing different causal chains, there is no such thing as a global past, present, and future.
5. The direction of causal order is from humans, as agents, to the measuring device to the system being measured. At least to this extent, humans are capable of asserting their will to choose which quantity they wish to measure, and are not forced into this decision by the system being measured. This defines a direction of causal flow. Specifically, the cause is the human's choice of the observable, and the effect is that the observable is measured, i.e. acquiring a single value, but not a specific value. In QM, humans only rarely have the power to choose a specific value (see section VI).
6. Humans, as agents, cannot be influenced by events at a later time, except by mental anticipation. Therefore humans can only experience the progression of causal chains forward in time.

Whereas the general public may object to principle #3, which defines some characteristics of time, this is non-controversial among the physics community, and is a natural deduction of special relativity. The simplest way to see that our perception of the flow of time is associated with causality is to look at any two events in classical special relativity. If these events are time-like, close enough in space and different enough in time so that signals can go from one to the other, they must be in the same time-sequence in all reference frames. This is consistent with principle #6, where the observer sees all causal chains going forward in time. If the two events are space-like, close enough in time and separated enough in space, so no signal can go from one to the other even at the speed of light, they appear in either time order, depending upon the velocity of the observer. Since the two events can't be in the same causal chain, there is no apparent flow between them, i.e. time doesn't have any flow, and the events don't have a unique time sequence. In QM, two space-like events appear to be causally connected, and this means non-locality. This paper will show that backward causation, causal chains moving backward in time, and which don't violate principles #5 or 6, can explain this non-locality consistent with special relativity.

Principle #2 is where I have received considerable objection among physicists actively interpreting QM. Basically these physicists either assume everything, including the future, already exists or that everything that can happen will happen, according to its own probability. Stephen Hawking, in Universe in a Nutshell[2], describes causal chains and raises the question: "Suppose there is a first event. What caused it?" He then confesses that many physicists, including him, want to avoid this question. On the next page he talks about casinos and rolling dice, which he compares to a universe experiencing multiple histories, each with their own probability. He follows this with a pictorial of Feynman's path integral, in which a particle takes every possible path between two points. Hawking is trying to eliminate causal chains.

Newton's first law states that a particle, with no force acting on it, will move at a constant velocity. This, in essence, is a causal chain, in which each point along the particle's well-defined path is an event in the causal chain. Newton's first law defines a causal chain, which allows us, in principle, to calculate the particle's location at each time precisely, based upon its earlier position and constant velocity. Hawking rejects this classical view, even in the retrospect of having detected precisely the space-time location of the particle at the beginning and end of its journey. Hawking claims that since the particle's velocity was never directly measured, it had many velocities, changing with time along its path and that a causal chain doesn't exist. Later in this paper we will identify interpreters of QM, who view the universe as a block universe, in which the future already exists.  Nothing is becoming, and relationships between realities in space-time are wrongly labeled as causal chains.  Causal chains, by definition, must have direction from cause to effect.  A simple relationship without direction between events cannot be considered a causal chain.

Another complaint against my model is that principles 5 and 6 lie well outside the accepted framework of physics.  If neutrality on these principles are enforced then all interpretations of QM should, at the very least, accommodate the possibility that principle 5 is true.  In fact none of the many interpretations, which assume a block universe, allow for principle 5.  Furthermore we could replace humans with computers, which are automated to choose the measurement configurations, either by random numbers or some other criteria.  Using arguments based upon the second law of thermodynamics we could argue that the processing of information is going from the computer to the measuring device, which defines a causal direction.  Principle 6 is justified based on human consciousness, by which we are aware of the past but not the future.  If arguments are presented that there are rare exceptions to principles 5 and 6, I will accommodate them by saying that these principles are not necessarily absolute, but rather are the normal progression of nature.

In our model causal chains, principle 2, are more than relationships and are actively changing the reality.  What is to become reality in the observer's future is ontologically partially different from the reality in the observer's present.  Here we are referring to the observer's time coordinate, without placing any restriction on the reality's time coordinate. The reality's time coordinate is inclusively future and past, meaning all of reality.  To understand this concept it is necessary to give up the idea that time flows.  The correct idea, that the flow is in the causal chains, is most clearly stated in principles #4 and 5.  Principle #6 reinforces the idea that causal chains have a unique direction, but this principle should not be interpreted to mean causal chains go only forward in time.  These ideas require modified word definitions.  Words, which we usually view as temporal words, actually refer to causal processes.  **Becoming** refers to the process moving along a causal chain.  **Until** may refer to the beginning of a causal chain, moving backward in time.  For example, events earlier in time of a particle's history may not be a reality **until** the experimenter chooses to measure a property of a particle later in time.  Similarly **before** and **after** refer to a position on a causal chain and not necessarily an earlier and later time respectively.

This discussion assumes that backward causation can occur, but it is not one of the basic principles, because it is too controversial.  Nothing in these six principles completely prohibits backward causation, however we will need to change the definition of an event, principle #1, to easily accommodate backward causation.  This change of the meaning of an event will be done after examining the objections to backward causation.  Those who still cling to a temporal meaning of the word **becoming** will object to the past being in a state of becoming.  For several reasons there appears to be an arrow of time always pointing into the future.  For example, the second law of thermodynamics states that entropy increases but doesn't decrease as time advances.  Others, who believe in a block universe, will also object to the concept of the past in a state of becoming, because nothing is in a state of becoming and even the future already exists.  Another problem with backward causation is the likelihood of illogical causal loops.  For example, it should be logically impossible to go back in time and kill one's own grandfather as a child.

To address these objections to backward causation we will use an analogy of a straight river with flowing water. The physical position along the riverbank represents time, with increasing time direction downstream. The flow of the water represents the causal direction. These two directions are generally the same – viewed on a large scale. But now suppose there are many small barriers in the river, each of which partially blocks the flow of water. Suppose also that the events downstream of the barriers can cause turbulence leading to eddy currents behind the barriers. The region of the eddy currents, behind these barriers, will exhibit backward causation; the flow of water (causality) will be reversed with respect to the riverbank (time). The eddy currents are affected by both the barrier (at earlier time) and events downstream at a later time, and the eddy currents flow both upstream (backward causation) and downstream (forward causation). In actuality the backward causation and forward causation may take the same path. Therefore the analogy requires two types of fluids moving through each other in opposite directions. This strange behavior is one reason we have to expand our definition of events, because a common chain, with the same events, is experiencing a causal flow in both directions. This becomes confusing if we consider the reality at a space-time location, i.e. an event, after only one of the causal chains has acted.

Using this analogy, the beginning of the river is the initial boundary condition of the universe and defined as a first cause. Other first causes, if they exist, can be thought of as rain drops falling on the river. It is the causal ordering of the universe, not the temporal ordering, which can explain the time asymmetric second law of thermodynamics (increase of entropy). For some reason, the initial boundary condition (first cause) of the universe appears to have been a very smooth, small-scale homogeneous state. This initial state had very low entropy in a gravitationally dominated universe where objects want to clump together. Therefore states of the universe further along in the causal order are found in a more probable state of higher entropy. The asymmetry arises from a low-entropy initial boundary condition on the universe, but no corresponding final boundary condition.

If we are proposing that an event A at a later time can cause an event B at an earlier, the issue of causal loops must be resolved. One can ask the question: what restrictions must be put on A and B to prevent causal loop paradoxes? The answer, which can be arrived at from first-principles, is simply that the backward causal effects on B may not be known by the rest of the universe, until after A occurs. If an outside measurement is attempted on B before A occurs, the measurement itself must interrupt the backward-causal process. Remarkably, this is exactly the sort of behavior seen on the quantum scale; measurement of quantum states is an inherently perturbative event. Quantum mechanics is exactly the sort of physics one should expect if backward causation is present on a micro-scale. Huw Price coined this concept with the words "inaccessible past", which is the past affected by future events, in his book <u>Time's Arrow and Archimedes' Point</u>[3].

With backward causation we must modify principle #1. In the standard view, in which the process of becoming flows with time, reality can be fixed at each four-dimensional space-time point. However, if the process of becoming is occurring at a frozen space-time point (no flow of time) then reality is also changing at this space-time point. Using principle #1 would mean that an event would also be changing. To make an event a fixed reality, rather than a process, suggests that we redefine it as the final reality at a space-time point.

Placing the adjective "final" in front of reality opens a new set of questions about reality. What is reality, if it is not final? Fortunately QM has exactly the right attributes to allow us to accommodate this question. The QM wavefunction or state vector, SV, is a probability distribution. Probability can refer to three things: lack of knowledge, the fractional part of a complete ensemble of events, or an unrealized potentiality. The majority, who interprets QM, would rule out the first option, and anyone who claims QM is complete would have to go with the third option. The third option claims that this unrealized potentiality is the full reality. There isn't anything else. Of course many interpreters claim QM isn't complete and start adding undetectable additions to nature, such as hidden variables, which don't have any direct evidence of existence. I find this approach very unsatisfying.

If the state vector, SV, is unrealized potentiality and also the full reality, then the reality itself is unrealized potentiality. If it is unrealized potentiality, then the potentiality must have the possibility of becoming realized. If we, as observers, view this potentiality from a temporal perspective and believe in causal chains, it would be natural to expect the realization of any potentiality to be caused by events later in time. Past events have not caused it, because it currently is not realized. Events in the observer's future bringing present or past potentiality to reality are defined as backward causation. An alternative to causation flowing backward in time would assume additional dimensions through which causation (process of becoming) flows. However it is usually the inclination of scientists to go with the simplest explanation first, which in this case is to limit our discussion to four dimensions and go with backward causation.

This interpretation of QM is easily applied to the double slit experiment, in which a stream of electrons is impinging, one at a time, on a barrier with two narrow parallel slits. Far behind the barrier the distribution of detected electrons on a screen resemble the standard interference pattern of wavelets emerging together from both slits (Young's double slit experiment). Many attempts of interpretation of this process go against the basic tenants held by the physics community. To have a whole electron pass through each slit simultaneously, without an accompanying positron, would violate lepton number conservation, which unlike energy conservation should not even momentarily be violated. To have part of an electron pass through each slit would violate our understanding of an electron. The interpretation most compatible with our understanding of nature is that the interference pattern is caused by probability (or propensity) wavelets emerging from both slits. To bring one of these propensity wavelets to reality would require a different detection system, which can identify the slit from which the electron emerged. Since the detection takes place at a later time than the electron passing through the slit, a causal chain beginning with the detection event must be going backward in time, bringing one of the propensity wavelets to reality. If the chosen detection system can't identify the slit from which the electron emerged then the wavelets from both slits remain as propensities, creating the interference pattern on the screen. In this case the earliest temporal event is the one that defines where the electron hits the screen. There would not be a causal chain defining the slit through which the electron passed.

To summarize our model, an event is not reality at a time T if all of its causes do not exist before time T. The event cannot be said to be 'real' until another time T', after which all of the event's causes have occurred. At this time T', the prior event can be said to have become a reality. The becoming at a later time T' of a reality at an earlier time T, gives a false impression that the concept of becoming is a temporal flow forward in time. This impression comes from principle #6, which states, using our analogy, that observers are always in the main current of water (causality) flowing downstream (forward in time).

Before continuing with the model of backward causation, it is important to clarify that this model is pure metaphysics, which is leaving unchanged the formalism and predictions of QM. However it claims QM is incomplete because QM ignores backward causation. Specifically at each moment in time there is a SV whose development in time is determined by the Schrodinger equation up to the collapse as viewed by the macroworld. The collapse of the SV is understood as the SV at earlier times being affected through backward causation of a macroscopic measurement on it. There is no modification of the time evolution of the SV as described by the Schrodinger equation. The Schrodinger equation does not describe or predict the collapse of the SV, and the mechanics of backward causation is solely limited to the process of the collapse. The proposal that the collapse causally moves backward in the time coordinate is an extension of the standard QM and addresses questions and processes, which standard QM does not address. There is absolutely no modification to the predictions of QM. The reason that QM, which describes only the forward time evolution of the SV, is always correct in predicting our observations is that we, as observers, only exist in the causal chains moving forward in time (principle #6). This means that principle #6 is essential for our model to be consistent with QM.

This model will interpret probability in QM as unrealized potentiality, which can only be brought to reality by future events. This interpretation requires the potential of backward causation for all QM systems. Principle #6 states that humans cannot be influenced by later events, and therefore we must conclude that QM is not universal. It is ironic that principle #6, which assures that QM always correctly predicts our observations, is also the principle, which insists that QM is not universal. This leads to several model dependent principles or definitions.

7. QM is not universal. It pertains only to systems, which can possibly be affected by future events (backward causation). Since humans cannot be so affected QM doesn't apply to whole humans. Similarly it probably doesn't apply to cats either (i.e. Schrodinger's cat), but this cannot be easily tested because of the concept of inaccessible past.
8. Definition: The realm, locus of all space-time points, where QM is valid and where causation has the potential to go in both time directions is defined as the microworld. The macroworld is the realm where causation is restricted to the forward in time direction.
9. Quantum measurement is an event, which is in both the microworld and macroworld (i.e. at their interface or boundary). Since this measurement is also in the macroworld it can never be subjected to effects from future events. However it can begin a causal chain going backward in time in the microworld, bringing to reality aspects of the temporal past, which were not previously a reality, i.e. past in a state of becoming.  This explains the measurement problem, i.e. the relationship between an observer and the microscopically observed object.  Furthermore, since the observer has free will to observe what he/she wishes (principle #5), this measurement event must be thought of as a "first cause" starting new causal chains in both time directions.

1. (modified) An event, b, is the final reality at a localized space-time region $\Delta x_b, \Delta y_b, \Delta z_b, \Delta t_b$.  In QM (microworld) the event is usually not point-like because of the Heisenberg uncertainty principle.  Final means that the reality is no longer subjected to change by future events.

The modified principle #1 is intended to be ambivalent concerning point-like reality.  It is not ruling out point-like reality, but rather leaving the door open for a more fuzzy reality.  If there isn't any point-like reality at a locus of space-time points, should we deny the existence of events at this location?  Our model doesn't have a general answer.  The next section on stationary states will demonstrate the lack of point-like reality when there are no causal chains.  Whether or not events in a causal chain are point-like is unclear.  As we apply this model to QM, many of the strange paradoxes and inconsistencies will be better understood, than in the past by other interpretations.

### III.    Stationary States

The ground state of hydrogen consists of a proton and an electron with zero orbital angular momentum.  This means that the proton and electron can only move in a radial direction towards or away from each other.  This state is known to be stationary, meaning that all observable properties are static, or unchanging with time.  Our model gives a clear understanding of why stationary states exist.  These states are stationary because they don't have any internal causal chains, or any events.  One reason we can conclude there are no causal chains is that the structure of the state in no way depends upon how it was formed nor on any past experience the hydrogen atom may have had.  Also, there are no events, which are final reality according to principle 1 (modified), because an experimenter in the future is free, in principle, to insert a probe into any nearby hydrogen atom to measure either the momentum or position of its electron.  These measurements bring to reality, what was previously not a reality.  Let us examine this in detail.

The Fourier transform of the ground state spatial wavefunction of the hydrogen atom is:

$$f(\mathbf{k}) = 2\,(2a_0^3)^{1/2} / [\pi(1 + a_0^2 k^2)^2] \qquad (1)$$

Where **k** is the wave number and $a_0$ is the Bohr radius.  The calculation of the probability of measuring a particular component of **k**, along an arbitrary z-axis is

$$P(k_z) = 2\,a_0 / [\pi(1 + a_0^2 k_z^2)^2] \qquad (2)$$

It is common to identify $\hbar\,\mathbf{k}$ as the momentum of the electron in the hydrogen ground state and $\hbar\,k_z$ as its component along the Z-axis.  To verify this distribution in the year 1937, x-rays were scattered off electrons in a large number of hydrogen atoms in their ground state[4]. If we define, **p**, as the momentum of the electron immediately prior to the x-ray scattering, standard kinematics gives the equation

$$2\mathbf{p} \cdot \Delta\mathbf{P} = \Delta P^2 + 2M\Delta E \qquad (3)$$

where $\Delta\mathbf{P}$ ($\Delta E$) is the change in the x-ray's momentum (energy). Letting $\Delta\mathbf{P}$ define the direction of the arbitrary Z axis, the calculated experimental distribution of $p_z$ values using the x-ray data and experimental equation (3), gave an identical result as the theoretical prediction of the hydrogen atom, equation (2), within experimental errors.

Let us discuss the meaning of $p_z$. It is not the momentum of the electron immediately after the x-ray scattering. Kinematics gives a different distribution of the post-scattering momentum. Nor can it be the momentum of the electron before the x-ray scattering because the hydrogen atom is in a stationary state. We know this both from theory and experiment. The QM state has no time dependence except for an overall phase angle, and both the mass and charge distributions are calculated to be static. The hydrogen atom has been examined extensively in the laboratory and all data is consistent with it being stationary. Both the mass and charge distributions are time independent. All ground state hydrogen atoms are indistinguishable at any instant in time, which also suggests a stationary state. However the definition of non-zero momentum necessitates the movement of mass (energy) from one spatial location to another. Of course several simultaneous movements can cancel each other out. However by definition, **p** is the total electron's momentum, and there is nothing to cancel out its mass and charge movement.

The standard Copenhagen Interpretation says that the electron doesn't have a momentum before the measurement. The measurement process brings to reality its momentum. In this interpretation the reality can only exist for an instant because after the x-ray scattering the electron has a different momentum. Many people interpreting QM today don't like the concept of measurement creating reality, and they reject the Copenhagen interpretation.

The Everett[5] multi-universes approach also allows measurement to modify reality. However it is not so much the creation of reality but rather the splitting off of reality into an infinite number of universes. If the electron simultaneously has all possible momenta, according to the theoretical distribution prior to measurement, then this would be consistent with a stationary state. Upon measurement, a specific value of $p_z$ becomes an exclusive reality in our universe and all other values of $p_z$ become a reality in other universes. This interpretation assumes all of the allowed values of momentum already exist and it is not being created, but only redistributed amongst multi-universes.

Having a momentum exist only for a point-like instant in time is nonsensical. Momentum, by definition, involves the spatial relocation of mass/energy over a finite time. If momentum only exists for a point-like instant at measurement there is no movement of mass/energy and therefore no momentum. To resolve this paradox would require time to be quantized into discrete steps of finite width, rather than being a continuous coordinate.

Wave-particle duality is also a pertinent issue here. QM describes the ground state of hydrogen as stationary standing waves of many different wavelengths without time dependence, equation 2. The human understanding of wave versus particle treats these concepts as ontologically different. A particle has a center-of-mass, whose spatial movement with time is described by its momentum.  Since a particle has only one center-of-mass there can be only one realized momentum describing its movement.  Such a realization occurs if the measurement of its momentum is made.  In contrast, a standing wave can be a superposition of waves of many wavelengths, $\lambda$.  The particle-wave duality comes from the de Broglie relationship:

$$p = h/\lambda = \hbar k \qquad (4)$$

Whereas a wave can have many simultaneous values of $\lambda$, a particle can have only one realized value of p.  This is true both conceptually and experimentally, and makes equation (4) nonsensical.  Anyone who claims an ontological equality in equation 4 is fooling himself or herself.  If we acknowledge that the reality of the internal structure of the ground state of hydrogen is unchanging over time, then any reality to equation (4) must come from the measurement process.  The measurement process must alter the reality, so that the measured reality, p, is related to the before measurement reality, $\lambda$.  Both the Copenhagen and Everett Interpretations assume that the measurement alters the reality, and our interpretation must make the same assumption.

In our model any change in reality must be part of a causal chain (principles 1 and 2).  Specifically the change, which occurs, must be identified with a causal flow rather than a temporal flow, as stated in principle 3.  When we refer to the words "before" and "after" measurement, we are not referring to the time coordinate of the hydrogen atom.  Rather these words refer to the time coordinate of the experimenter.  The experimenter, by scattering an x-ray off an electron, is creating a causal chain propagating into the past of the hydrogen atom.  This process brings to reality the momentum $p_z$ into the electron's temporal past, but not the experimenter's temporal past.  Our model is identical with the Copenhagen interpretation, that the measurement creates reality.  It is different from the Copenhagen interpretation in that it allows the electron's past reality also to change.  In our model the potentiality in the past is identified with the wave nature of the electron, whereas the past reality or final events, which are part of a causal chain caused by the future measurement, are identified with the particle nature of the electron.  The equal sign in equation 4 means that what is potentiality may become reality.

In physics there are conservation laws, in which certain quantities are conserved throughout time.  These quantities must always be a reality independent of events, causation, and causal chains.  Some of the important conserved quantities existing in the ground state of hydrogen are lepton number, electric charge, energy and angular momentum.  The conserved quantities completely determine the properties of the ground state of hydrogen.  Initial or final boundary conditions, due to the surrounding environment, have absolutely no effect for the state.  The hydrogen atom has one electron, which must exist, although it has no momentum in space-time.  Conservation of energy and angular momentum specify exactly where the electron, relative to the proton, has a propensity to exist.  Since energy conservation can be violated for short periods of time, this propensity distribution has exponential fall-off at large distances from the proton.  Since electric charge is conserved and must be present, it is distributed around the proton as a spherical cloud, with a spatial distribution given by the distribution of the propensity of the electron.

Just as the electron exists but doesn't have a value for its momentum or position, because there is no causal chain, so also energy exists but is not realized as either potential or kinetic energy.  In this sense there is no final reality, but only a propensity for reality. This raises the philosophical question, how can something exist without having a value, or how can energy exist without having a form? Using ideas borrowed from Aristotle[6], we call some of the substance of the universe eternal, or "essential". However some of the elements present in things are "accidental", resulting from cause and effect, which represent change. Using these ideas, we say that the hydrogen atom is not normally subjected to any causal chains and therefore is not undergoing change. However, as long as it is left alone all of its properties, determined by conservation laws, are essential. A reasonable belief, based on causality, is that the electron's future momentum currently lacks reality. Time symmetry suggests a similar property for the past. The laws of physics, as they pertain to the hydrogen atom, are completely time-symmetric, and momentum, and anything else subjected to unrealized causal chains, lacks reality in the past as well as the future. Time does not flow, and causal chains, if they are lacking, must be absent in both time directions for stationary states. There should be no distinction between past and future.

This interpretation of the ground state of hydrogen assumes that it is immune to causal chains. Let us look at this more closely. Suppose we insert a probe and measure the momentum of the electron. This will initiate causal chains going both forward and backward in time, along which the electron acquires a momentum. Here we are not interested in the magnitude of the momentum, since the magnitude varies continuously, dependent on the unknown position of the electron. Rather we are interested in the momentum's direction, which is conserved. The direction of the electron's momentum in the rest frame of the atom can be determined either by measuring the position or momentum of the electron relative to the proton. For simplicity we will assume the uncertainty in direction is negligible. With the direction defined, the ground state, in its inaccessible past, will become a thin line centered at the proton. This replaces its usual spherically symmetric shape. The direction of the measured momentum defines the orientation of the line. In this configuration the electron is actually moving, because of the reality of the causal chain, and it hits the proton head-on. The collision of the electron with the proton results in a probabilistic recoil angular distribution. This probability, inherent in the collision, causes the hydrogen atom to revert completely to a spherical shape after only a few collisions, going into the past. Only the immediate past is affected by the measurement and this past is probably "inaccessible", i.e. can never be observed. Similarly only a brief future of the atom can be affected by measurement, assuming it stays in its ground state. We can safely conclude that the atom is immune to both initial and final boundary conditions in that there are no long lasting causal chains.

As stated earlier the electron has a propensity distribution, which has an asymptotic exponential fall-off with distance from the proton. This distribution could conceivably be examined using an electron tunneling microscope probe. Although there is zero uncertainty in the energy of the hydrogen ground state, energy conservation can be violated briefly. This allows the probe to capture the electron at a distance from the proton where the coulomb potential energy with the proton is greater than the total energy. The probe provides enough negative potential energy to allow the realization of the propensity for the electron to exist at such a great distance. Most interpretations of QM would call this tunneling, where the electron is initially inside the coulomb barrier, tunnels through, and appears on the other side. In our interpretation there is no causal chain going backward in time from this detection event. There is no motion through the barrier, which is impossible because of the lack of kinetic energy inside the barrier. Rather the electron simply has a propensity to exist where it is detected, and this has been brought to reality by the probe. The distribution of the electron's charge is based upon propensity rather than the reality of events, because the ground state doesn't have any events. There is no causal chain of events backward in time because the electron can't experience real events inside the barrier, and doesn't have momentum.

The early founders of QM struggled to understand stationary states and finally gave up. Physicists, for the most part have given up trying to understand these states. The Copenhagen interpretation doesn't try to give a realistic view of these states. Every realistic dynamic interpretation of these states assumes the electron is continuously accelerating in the hydrogen ground state. Therefore all dynamic interpretations create a contradiction with electro-magnetic theory. E&M theory has a fundamental principle that any charge experiencing acceleration must be emitting and/or absorbing radiation during its acceleration. Our interpretation treats quantum probability, associated with stationary states, as a propensity or potentiality without real events. Nearly every other interpretation of stationary states treats the quantum probability as either a lack of knowledge of an existing reality or as a distribution, in which all possibilities are realized. Either of these alternative views of probability assumes the electron has position(s), momentum (momenta) and is moving and accelerating. Our model solves this contradiction with electromagnetic theory and explains the particle-wave duality. The key to understanding this is to accept the idea that a particle can exist without having a definable momentum and location. If it has momentum, this property would be a causal chain. Alternatively, without causal chains there aren't events, which means a lack of final reality in space-time.

## IV. Non-Stationary Atoms

In QM a non-stationary bound state has a SV, which is not an eigenstate of the Hamiltonian. This can happen because of some external perturbation and is usually transitory in nature. The best-studied non-stationary states are in Rydberg atoms. This refers in general to atoms with an electron weakly bound in a very large orbit. The motion of an electron in its non-stationary Rydberg orbit can be observed in various ways. For example[7] a short optical pulse excites an electron into a superposition of Rydberg orbits forming a small radial wave packet. The wave packet moves classically in and out from the ionic core. Only if the electron is near the ionic core will photo ionization by visible light occur. A collection of such Rydberg atoms are formed identically by the same optical pulse and their behavior is monitored by photo ionization. Intensity peaks in this ionization are observed at times after the optical pulse, which are integral multiples of the classical round-trip time of the electron moving in its orbit. This confirms that the electron's movement approximates this classical orbit.

Whereas experimenters can obtain clear evidence of motion in a non-stationary state, it is impossible to detect by non-perturbative means any motion or change in a stationary state. I am unaware of any interpretation of QM, which gives an ontological explanation of the difference between these two types of states. Of course most physicists are not interested in this since the math is clear in its distinction between stationary and non-stationary states. It can be argued that definitive mathematical statements speak for themselves and don't need an interpretation. We would like to argue that many mathematical statements have a deeper meaning, which goes beyond the mathematics. In our model a non-stationary state has a causal chain, which is bringing to reality the motion of the electron. This is what makes it fundamentally different from a stationary state.

A relatively new research area is the coherent control of quantum systems. A paper[8] titled "Quantum Physics Under Control" gives an excellent summary. Most commonly this involves tailored shaped pulsed laser fields manipulating quantum mechanical processes. Much of this work is empirical in that making small changes by trial and error in the control field maximizes the desired outcome. This is done through either real-time feedback on a single system or by learning control on a sequential set of identically prepared systems. In all cases these involve non-stationary processes experiencing cause-effect, i.e. causal chains, and the causal chains progress from the apparatus and ultimately from the experimenter (principle 5). Some of the controlled systems are vibrational or rotational excitations of molecules or chemical reactive processes.

The measurement of the momentum of the electron in the ground state of hydrogen, discussed in the previous section, converts the hydrogen atom to a non-stationary state. During a brief past before its momentum measurement, the electron, having acquired a momentum, is moving, accelerating, and giving off electro-magnetic radiation. Energy non-conservation, as given by the Heisenberg Uncertainty Principle, is crucial for this process. The electron must acquire some extra energy through the measurement process, and this energy can be radiated off earlier in time. The total energy of the system can fluctuate over a brief period of time.

### V.     Measurement

One of the major components of any interpretation of QM is to define the mechanism of a measurement on a quantum system. Many consider this the most outstanding problem in the interpretation of QM. In general this involves a potentially observable property of a quantum system serving a critical role in the interaction of that system with some other object. If the interaction doesn't force the quantum system into an eigenstate of the operator, which represents that observable, then no measurement is made. Instead both the quantum system and the object with which it interacts become entangled into a superposition of two or more eigenstates with two or more different values of the observable. This would be the situation for the hypothetical Schrodinger cat, which is both alive and dead simultaneously.

If alternatively the quantum state suddenly changes from a superposition of two or more eigenstates to a single pure eigenstate a measurement is taking place. It is commonly recognized that the major difference between a measurement and a non-measurement is that the former is irreversible, i.e. the result finalized, and the latter reversible process is not finalized. In the model of backward causation this property is clearly understood. The word "finalized" means a potentiality becoming a reality. In the case of a non-measurement we are in the realm where causation is still free to act on the current state from some future event, which has not yet occurred. This is what non-finalized means. The system is entirely in the microworld where causation is free to move in both time directions. In contrast a measurement represents the interaction of a quantum system with the macroworld where causation progresses only forward in time. Since the causal chain into the future of the macroworld cannot reverse itself, the process is irreversible and cannot be changed.

There have been two common conjectures, consciousness and multi-universes, about what causes the difference between a measurement and a non-measurement. The first of these conjectures suggests that unless human consciousness is involved there is no reduction of the quantum state to a pure eigenstate of the observable. The Everett[7] multi-universe (multi-world) approach accounts for the difference between measurement and non-measurement by having all possible outcomes branch off into other universes for all measurements. If no branching occurs there isn't a measurement. Both conjectures are based upon ignorance and neither has a definable mechanism. In contrast, the causal chain argument, based upon principles outside of QM, not only demands the difference between measurement and non-measurement, but also clearly explains the reasons for these differences. A measurement is recorded as an event in the macroworld, which is final because causation in the macroworld only goes forward in time. The only thing lacking in the causal argument is the identification of the boundary between the microworld and the macroworld.

Our interpretation of quantum mechanics takes a realistic view of measurement. Specifically, if a scientifically valid measurement is made on an object, its registered value is a true indication of the properties of the object both immediately after and prior to its measurement. However, this model includes backward causation in which measurement may actually bring to reality the measured property earlier in time. Because of this active role of backward causation counterfactual arguments involving possible measurements (i.e. "what-if" measurements) are susceptible to error. The error is that the backward causation is ignored. The counterfactual argument assumes that the system is in the same state before the measurement is made regardless of whether or not the measurement is actually made. The argument makes the reasonable, but wrong, assumption that if the outcome of the what-if measurement can be predicted with 100 percent certainty, then it is a reality, whether or not the measurement is made. Such counterfactual arguments are common and lead to apparent paradoxes and inconsistencies.

One of these, Hardy's paradox[9], will be discussed next in this paper. Hardy says, "If we can predict with certainty (i.e. with probability equal to 1) the result of measuring a physical quantity, then there exists an element of reality corresponding to this physical quantity and having a value equal to the predicted measurement result." Hardy borrowed this statement from the original paper by Einstein, Podolsky, and Rosen[10]. Here Hardy is claiming that the reality is independent of the measurement, i.e. the measurement doesn't actually have to be performed for the reality to exist. Another possible problem with what-if measurements is that they may be on two non-commuting variables. In this case one measurement will block the causal effects of the other measurement, so that only one measurement is meaningful.

### VI.     Hardy's Paradox

Hardy states, "If elements of reality corresponding to Lorentz-invariant observables are themselves Lorentz-invariant realistic interpretations of quantum mechanics are not possible." He is saying that a realistic interpretation of QM is not consistent with special relativity. He bases this conclusion on a gedanken experiment involving a single positron and a single electron each moving through separate interferometer systems, which overlap. Rather than describing the specifics of the experiment and his analysis of it, I will describe in more general terms how his argument develops.

Detectors are placed at the ends of each interferometer, so that one detects the positron and the other detects the electron. With his detector arrangement, and the experiment repeated many times, the positron and electron detectors will together get positive readings 1/16 of the time. Now, these detectors are so far apart from each other that their measurements are space-like events. This means that the time ordering of the measuring events can be different in different inertial frames. Hardy considers the inertial frame where the positron detection takes place first. With a positive reading in the detector QM predicts with 100% the path, which the electron takes. Next Hardy considers another inertial frame in which the electron detection occurs first. With a positive reading of the electron QM predicts with 100% certainty the path, which the positron takes. Hardy then points out that these two properties are mutually inconsistent. Specifically, if the electron and positron were both on their predicted paths they would meet and annihilate, and neither would be detected. The prediction involves conclusions based on observations in two different inertial frames, so that Lorentz invariance is assumed. Since the prediction, made with 100% certainty is wrong, a realistic Lorentz-invariant interpretation of QM is impossible.

Hardy's paradox is a subtle counterfactual argument, which seems very reasonable because he is simply viewing the system from two different inertial frames. Lorentz-invariance requires the reality of the system to be the same as viewed in different inertial frames. If the two realities are inconsistent then reality is not Lorentz-invariant. The counterfactual part of the argument is veiled because he never explicitly discusses the verification of the properties of the system, which are predicted with 100% certainty. Let us say that we want to verify these specific properties. We consider first the property that is predicted from the detection of the positron. To verify the path of the electron its detection system would have to be totally modified before the electron is detected. Similarly to verify positron's path that is predicted by detection of the electron the positron detection arrangement would have to be significantly altered before the positron is detected. Here I am assuming that QM gives a completely accurate prediction of measurement, and either verification would be successful if the detection system was altered to verify the prediction. Specifically, the electron, or positron with a different detector arrangement, would always be found on the predicted path, with the experiment repeated many times.

Conceptually, one may argue that the necessity of altering the detection system doesn't invalidate Hardy's argument because we are talking about properties of the system which extend back earlier in time, before either measurement is performed. I would like to turn Hardy's so-called inconsistency argument into a proof of backward causation. In order to have a realistic Lorentz-invariant interpretation of QM, there must be backward causation to invalidate Hardy's argument. Other solutions (see below) to Hardy's paradox have serious problems.

In our model each detector has a causal effect on the system moving backward in time. Furthermore, neither causal effect is deterministic enough to block the causal flow of the other. However, the two causal effects interfere with each other, so that the two effects don't simply add to each other. The ordering of the two measurements has no effect on reality and the causation of both detectors must be included. If one applies the two measurements as final boundary conditions on the quantum system, one finds that these so-called earlier properties of the system are ambiguous and can't be deduced with 100% certainty. There are, in fact, several possible realistic descriptions of the earlier states of the electron and positron, which are totally consistent with the positive reading in both the electron and positron detectors. Whereas the what-if measurements, which are not set up by Hardy, would give the quantum system a deterministic outcome, the actual measurements, which he considers, leave the electron/positron state ambiguous. Hardy is wrong in concluding that a realistic interpretation of quantum mechanics is not possible.

The following is a quote[11] from a peer review of my paper: "Hardy's paradox was resolved before in the framework of time symmetric approach, Vaidman[12], and I do not see that what is proposed by the author is better… it (referring to my model) presents a radically new theory…" This goes to the very essence of the disagreement between my so-called radical model and many of the physicists who deal with these issues. These physicists are rejecting principle #2 in the introduction and are favoring a block universe worldview. Vaidman uses the following equation, the so-called ABL rule, for calculating probabilities for results of an intermediate measurement performed on a pre- and post-selected system.

$$prob(A = a_n) = \frac{|\langle \psi_2 | \hat{P}_A = a_n | \psi_1 \rangle|^2}{\sum_k |\langle \psi_2 | \hat{P}_A = a_k | \psi_1 \rangle|^2} \tag{5}$$

Where $\hat{P}_A$ is a hypothetical measurement in the inaccessible past at a time between the initial prepared state $\Psi_1$ and the final measured state $\Psi_2$. The denominator simply normalizes the probability by summing the probabilities for all possible values of A in the inaccessible past. If alternatively the sum was over all possible unknown final states and $A = a_n$ was the known measurement, the revised expression would be the standard QM probability for obtaining a yet-to-be final state $\Psi_2$. In the Hardy paradox the final state, $\Psi_2$, is measured and known, but not the intermediate value of A. Vaidman describes the probability, equation 5, that $A = a_n$ as "elements of reality defined by both prediction and retrodiction." The word "define" is interpreted as a passive word, simply describing the situation. Specifically, there is no causal relationship between the post-selected state $\Psi_2$ and the measurement $\hat{P}_A$, because we are in a block universe where everything already exists.

There are serious problems with the ABL rule. Consider using this rule on the hydrogen ground state (section III.) in which the momentum of the electron is measured. $\Psi_2$ is the final measured state of the electron, and $\hat{P}_A$ would be a hypothetical measurement of its momentum earlier in time. A blind application of the ABL rule would give the nonsensical answer that the electron has plus or minus the spatial direction of the measured momentum, within the uncertainty of the measurement, for earlier times. Furthermore, the electron would have a momentum whether or not the measurement of $\Psi_2$ was made since it is assumed there is no cause-effect going backward in time. Vaidman would have to reject the application of the ABL rule to stationary states, because he undoubtedly believes these states truly are stationary and don't have any net momentum. This brings up the question of when the ABL rule is valid.

The ABL rule assumes the post-selected state must already be a reality before the final measurement is made. The final measurement has no cause-effect on the quantum system at the instant the measurement is made or into the past. This is the desire of many physicists to describe objective reality independently of measurement. This same idea also applies to the hypothetical measurement, $\hat{P}_A$, in the inaccessible past. In our model, we are forbidden from even conceptualizing a measurement in the inaccessible past because of the basic tenet of QM, taught in most QM courses, that the observer affects the observed. The ABL rule cannot logically accommodate causal chains because the cause-effect of both $\hat{P}_A$ and the final measurement giving $\Psi_2$ could lead to illogical causal loops. In contrast to Vaidman, Hardy uses the standard mathematical procedures of QM in which an observation affects the quantum state. The reality of the quantum system being observed, according to the mathematics in Hardy's paper, changes the probability distributions. If this change is ontological, then the order in which the positron and electron are detected is important. If the change is epistemic, as assumed by Vaidman, then the order of measurement has no effect on reality. In our model the change is ontological, but because of backward causation the order of the measurement is unimportant.

### VII. Probabilistic Causation

An essential property of QM, which allows for backward causation, is its stochastic nature. This means that causes are not sufficient conditions for their effects. This type of causation is called probabilistic causation[13], and is essential for understanding the EPR paradox in the next section. In our model an event in the inaccessible past of a quantum system will experience causal chains from opposite time directions. If either causal chain were deterministic the other causal chain would be blocked off from acting. Let A and B represent two events, which are causally related. Probability causation satisfies the following relation:

$$P(B|A) = P(A \& B)/P(A) \neq P(B) \qquad (6a)$$

Here $P(B|A)$ is the conditional probability that event B occurs given event A. If event A enhances the probability of event B we would have $P(B|A)$ larger than $P(B)$, but one should also include in the definition of causation event A inhibiting event B. Therefore the inequality gives a more inclusive definition. One problem with probability causation is thought to be its symmetry:

$$P(A|B)/P(A) = P(B|A)/P(B) \qquad (7)$$

This means that the occurrence of B changes the probability of the occurrence of A by the same fractional amount that the occurrence of A changes the probability of the occurrence of B. Naively with causal flow from one event called the 'cause' to the other event called the 'effect' one would want the causal relationship to be asymmetric. In my model symmetry between events is the preferred attribute because the causal flow moves both forward and backward in time. By maintaining this symmetry the time symmetry in the microworld is honored. Of course, if there is complete symmetry then one can question the whole notion of causal flow existing. To answer the question of causal flow in the affirmative we will consider the types of measurements to be chosen independently by two different individuals for A and B respectively. Then, according to principle #5 there must be two causal chains going in opposite directions between events A and B. Consider two spin ½ (in units of h/2π) particles initially prepared in a so-called singlet state in which their spins are anti-parallel. If one particle is measured to have spin +½ along any arbitrary axis then the other particle will be found with certainty to have spin projection −½ along the same axis. If these two particles move apart from each other undisturbed in opposite directions, maintaining this anticorrelation, they are called entangled. This entanglement can remain over large distances and is an example of the non-locality in quantum mechanics. These two measurements A and B will be applicable to our probability causation model. Let event A be a measurement of the spin projection of particle 1 along an axis, x, which is freely chosen by an experimenter. Suppose the event A measurement obtains the value −½. Let event B be the measurement of the spin projection of the other particle, 2, along another arbitrarily chosen axis, χ, selected independently by another experimenter. Suppose the event B measurement value is also −½. The probability of each measurement result, by itself, is 0.5 because there are two equally allowed spin projections, ± ½ for any axis.

$$P(A) = P(B) = 0.5 \qquad (8)$$

[note: decimal notation is used for values of probabilities to distinguish from spin projections]. Quantum mechanics interprets each measurement result as an eigenvalue of the measured observable with an eigenstate describing the two-particle spin state. The spin eigenstates for measurements A and B are:

$$\psi_A = \phi_{1,x}(-½) \, \phi_{2,x}(+½)$$
$$\psi_B = \phi_{1,\chi}(+½) \, \phi_{2,\chi}(-½) \qquad (9)$$

where $\phi$ is the single-particle spin function, the subscripts give the projection axes, the number 1 labels the particle at event A, number 2 labels the particle at event B, and ±½ are the spin projections. Mathematically each measurement selects a spin function for both particles. To help interpret whether this is real and causal, we need to relate these functions to actual experiments. If many measurements of this type are performed on identical systems, the probability distributions can be obtained, not only for this pair of events, but for numerous pairs of events. Since we want to keep the choice of axes independent, we must have measurements for many different, and randomly chosen axes. However, all probabilities listed below are for the sub-sample of events where axis x is freely chosen for event A and axis χ is chosen independently for event B. Events A and B are also distinguished by their widely separated spatial location in the lab. The probability of event A occurring (spin projection -½), given event B (also spin projection -½) is the same as the probability of event B occurring, given event A.

$$P(A|B) = P(B|A) = \langle \psi_A | \psi_B \rangle^2 \qquad (10)$$

Mathematically this quantum mechanical probability only depends upon the specific eigenvalue, ± ½, for each measurement and the angle between the two axes. Generally this probability deviates from 0.5, which is the probability of each event occurring in isolation (eqn. 8), and therefore the condition for probability causation (eqn. 6a) is satisfied. This suggests but doesn't guarantee that event A has a probabilistic causal effect on event B. Furthermore, the following relationship is also satisfied:

$$P(A | B) = P(A \& B)/P(B) \neq P(A) \tag{6b}$$

which suggests that event B has a probabilistic causal effect on event A, and demonstrates the symmetry between the two events. In order for the conditions 6a and/or 6b to be valid criteria for probabilistic causation between events A and B, we must rule out a spurious correlation from other causes. Suppose there is a C satisfying the relationship $P(B|A\&C) = P(B|C)$. In this case A is probabilistically irrelevant to B. Using the terminology of Hans Reichenbach[13], C is said to "screen A off" from B. There is a voluminous literature on the EPR paradox examining this question and no one has been able to even conceptualize a plausible C, which can screen events A and B off from each other. This is particularly true if we give the experimenters the free choice to choose the axes x and χ independently from each other, and we analyze different subsets of events, each with a different combination of axes. Furthermore, the spins of the two particles must, according to the definition of entanglement, be left undisturbed until measurements A and B are performed.

### VIII. Einstein, Podolsky, Rosen Paradox

We have examined the EPR paradox in the previous section on probabilistic causation, and this section will continue the discussion in more depth. QM is a non-local theory, meaning that events, which are simultaneous measurements, A and B, on two so-called "entangled" objects separated in space, are correlated as in equation 6a or 6b so that they appear causally related. Furthermore, it seems impossible to explain this correlation by the common past of the two spatially separated objects. QM seems to demand that causal effects travel between these events at speeds faster than light. However, this violates Einstein's Special Relativity. It should be emphasized that the mathematics of QM is not inconsistent with Special Relativity, but rather most metaphysical interpretations of QM are inconsistent with SR. Interpretations, which claim to be consistent with SR, may ignore the free will of the two observers, principle #5. Any proper interpretation must assume that two different observers, independently of each other, freely choose the type of measurements, A and B. The acceptable way, in my opinion, to avoid this non-locality problem is to allow for backward causation. We assume the particles haven't experienced any decoherence, loss of entanglement, through interaction with the rest of the universe, even though they have separated by a great distance. Causal effects of each measurement travel backwards in time to the common past of the two particles and forwards in time to the other measurement, at speeds less than the speed of light. The elapsed times for each part of the journey can cancel, or even give negative times. Therefore causal effects can travel between these correlated events in zero or negative time. QM has an important attribute making this interpretation completely acceptable. The indeterminism in QM prevents these causal effects from carrying useful signals, thereby preventing superluminal signaling.

Because space-like events can appear in either time order, cause-effect must also be able to go in both time directions between these events. In the EPR paradox let us make simultaneous measurements on particles in two separated regions 1 and 2 in the normal lab frame, labeled S', where the two particles are traveling at equal speeds, v' and distances, x' in opposite directions. Backward causation allows causal effects to travel great distances in zero time or perhaps backward in time. In the S' frame the effect of one measurement travels backward in time by the amount x'/v' to the point of contact and then forward in time by the amount x'/v' to the other measurement site. These times cancel and the effect takes zero time to travel from one measurement to the other. By using a Lorentz transformation we can calculate the time between these two measuring events in another inertial frame, S in which S' moves at velocity u. With $t'_1 = t'_2$ we get $t_1 = t_2 - 2\gamma\beta x'/c$ where $\beta = u/c$ and $\gamma = 1/\sqrt{1-\beta^2}$. We could equally well calculate these times by identifying one of these measurements, perhaps in region 2, as the cause, which affects possible outcomes of the other measurement in region 1. Picturing this cause traveling backward in time with one of the particles, to the point of contact and then forward in time with the other particle we can calculate the time between the measuring events in any inertial frame. Using the Lorentz transformation of velocity and spatial contraction as needed we subtract the time of motion during the backward causation stage from the time of motion during the forward causation stage. This gives the exact same times for the two measurements as in the straightforward Lorentz transformation:

$$v_1 = \frac{v' - u}{1 - v'u/c^2}$$

$$v_2 = \frac{v' + u}{1 + v'u/c^2}$$

$$t_1 - t_2 = \frac{x'/\gamma}{v_1 + u} - \frac{x'/\gamma}{v_2 - u} = -2\gamma\beta x'/c \qquad (11)$$

The effect of a measurement in region 2 arrives at region 1 simultaneous with the measurement at 1 and at an earlier time than the cause at region 2 in the S frame. That we got the same value for ($t_1 - t_2$) in two different ways demonstrates that SR is self-consistent, but it also vividly emphasizes that the dynamics of this problem can be treated within the SR framework. This is important to demonstrate because many have doubted the EPR paradox could be understood using SR. Notice that the time between measurements in the S frame only depends upon the velocity of the frame, u, and the spatial separation between the measurements. We could just as easily have chosen the measurement in region 1 to be the cause, have its effect travel backwards with particle 1, and then forward with particle 2. This view would give us the exact same time relationship between $t_1$ and $t_2$. Both measurements, in region 1 and 2, are causes which affect the two-particle SV in the past, leading to a double reduction (see below).

Backward causation gives a causal symmetry between the two measurements, which restores time symmetry. Without backward causation this symmetry is lost.

Having causality flow in both directions between space-like events appears to violate the unique causal order principle. The essence of this principle is retained because each measurement acts separately (i.e. multiple causes) on the SV, leading to a double reduction. Both the stochastic nature of QM and the Heisenberg uncertainty principle allow causal effects to pass through each other without contradiction. If a spin measurement in region 1 were allowed to totally determine the spin of the particle there would be no room for a spin measurement in region 2 to have any causal effect on the spin direction. QM gives enough free play so that both measurements always can have a casual effect on the SV, without any contradiction. This free play is manifested in the fact that any eigenstate of measurement A is not orthogonal to any eigenstate of measurement B. This means one measurement can not exclude the other measurement.

Assume the measurement in region 1 is along axis x and measures a spin down which is denoted by $-\frac{1}{2}$. With the restriction that the spins are anti parallel, particle 2 has spin up, denoted by $+\frac{1}{2}$, so that the SV at earlier time is reduced by measurement 1 to $\psi_A$ (equation 9). Similarly the measurement in region 2 can be along another axis, $\chi$, and possibly also obtain a spin down. The effect of this measurement is to produce the spin state $\psi_B$ (equation 9) at earlier time. Causally before a measurement occurs on either particle, particle 2's density matrix has its spin orientation evenly distributed in all directions. In the S reference frame where the measurement at $t_1$ occurs before the measurement at $t_2$, the causal order has particle 2 reduced first to $\phi_{2,x}(+\frac{1}{2})$ and later to a double reduction giving both $\phi_{2,x}(+\frac{1}{2})$ and $\phi_{2,\chi}(-\frac{1}{2})$. In an inertial frame, S", in which S' moves in the opposite direction, particle 2's SV first gets reduced to $\phi_{2,\chi}(-\frac{1}{2})$ followed by the double reduction $\phi_{2,x}(+\frac{1}{2})$ and $\phi_{2,\chi}(-\frac{1}{2})$. Each reduction becomes accessible to the macroworld at the time of each measurement. At all points along the trajectory, with each point representing a different earlier time, the SV appears suddenly reduced. The word suddenly refers to the current time of measurement and not the time represented by the wavefunction in the inaccessible past.

Causal processes are correctly associated with the time coordinate. This is clearly seen in SR and the limiting speed of light. If casual processes could instantaneously traverse space, they could be associated solely with the spatial coordinates. In the EPR paradox it appears as if the effects of a measurement don't involve movement through time, because the effects appear instantaneously at widely different spatial locations. This apparent lack of time progression along a causal chain is a misconception. Any two entangled particles have a common past, and this common past involves all causal chains connecting the particles. Causality must move through time. In the macroworld it is restricted to moving forward in time, and this leads to the misconception of time being defined by change.

In QM there is no such thing as a doubly reduced wavefunction. Specifically, QM does not allow a SV to have spin down along the x-axis and spin up along the $\chi$ axis with 100% certainty. The two measurements of spin along the x and $\chi$ axes do not commute and therefore cannot be simultaneously realized in standard QM. In our new model, the effects of each measurement approach each point along the trajectory from opposite time directions. There is no need for them to commute and the SV in the inaccessible past is doubly reduced. Only when one looks at times after the measurements do the effects flow in the same time direction. Since the measurements do not commute, each particle's SV at later times is only affected by the measurement on it and not the measurement on the other particle, thereby giving a result consistent with standard QM. One important feature of the double reduction of the wavefunction in the inaccessible past is its independence of the time ordering of the measurements. This is a requirement of Lorentz invariance because the time ordering of measurements is frame dependent. The past, which is converted from a propensity to a realization, should be the same in all inertial frames after all measurements are completed.

It is instructive to look at Griffiths' discussion[14] of quantum incompatibility. Griffiths considers the two-dimensional Hilbert space of a spin state of a spin half particle. He considers the spin state at one moment in time occupying one subspace of his Hilbert space. Logical inconsistencies are shown to occur when the subspaces are represented along different axes. For example, the subspaces using the x-axis are $S_x = +$ ½ or $-$ ½ (in units of h/2π) or alternatively $S_\chi = +$ ½ or $-$ ½ when using the χ-axis. He demonstrates logical inconsistencies between SVs occupying a single subspace in each representation of the Hilbert space and concludes that logical statements involving two representations of the Hilbert space are meaningless.

The backward causation model avoids these logical inconsistencies. Before the final boundary condition of a SV is realized by a macroscopic measurement, the SV represents a history of propensities. With backward causation it is more appropriate to view probabilities in QM ontologically, not as incomplete knowledge, but instead as propensities (dispositions or potentialities), which have not yet been subjected to the causal flow of post selection. Propensities, by definition, are distributive functions over two or more subspaces of the Hilbert space and are not restricted to a single subspace. If the SV is restricted to a single subspace, through either initial or final boundary conditions, then the boundary condition specifies the correct representation of the Hilbert space. All alternative representations, in which the SV is in a single subspace, are invalid. These restrictions avoid the quantum incompatibility presented by Griffiths. However, in our example of the EPR paradox, the SV is doubly reduced to a single subspace in two representations, x and χ. This special case only exists in the inaccessible past, which is not open to future boundary conditions, and is not treated by standard QM. The final spin state in the EPR paradox must be doubly reduced to be Lorentz-invariant and this must occur through backward causation to avoid the problem of quantum incompatibility.

Hardy[9] wrote down a QM wavefunction, which was reduced by a future measurement, and showed that different what-if measurements on this wavefunction would be logically inconsistent. These what-if experiments would identify with 100% predictability which path the electron or positron took through the interferometer. They were what-if experiments because the equipment was not set up to identify the path. This should be a warning that any attempt to deduce properties of the inaccessible past, based upon conceptualized experiments is fraught with error, because it violates the principle that this past is truly inaccessible.

A criticism of my model of backward causation is that there is no mathematical model describing it. However any attempt to construct such a model would be prone to error, without means of identifying the error. An attempt to write down a wavefunction for the inaccessible past, with the final boundary conditions, invites the same mistake

Hardy made. Whenever a person studies an equation, he/she likes to conceptualize it. How can we formulate a model of the inaccessible past without conceptualizing a what-if measurement on it? Any real measurement is a boundary condition on the equation itself. The equation is impossible to solve without knowing the boundary condition. My conclusion that the spin wavefunction is doubly reduced must be held tentatively. This conclusion was based on the assumption that the effects of two non-commuting measurements can both be realized if the causal chains come from opposite time directions. However, there is no foolproof way to test this or even know what it means without conceptualizing spin measurements on the electron in its inaccessible past. Basically I am arguing that it is unreasonable to demand a mathematical formulation of backward causation. Since the equations, without boundary conditions, don't have unique solutions, the value of a mathematical formulation is mostly limited to demanding logical consistency.

The properties of the inaccessible past are evolving separately from the progression of its own time coordinate. To represent this dynamic change we would need to introduce a new parameter, the causal order parameter. But such a representation would be hard to visualize because we can only picture change with time. Furthermore the causal order parameter is not a new dimension similar to space and time, but rather is connected with our four-dimensional space and time. Because it is connected with time, I would prefer to describe the microworld using a step function, $u_s(t'-t_b)$, which is equal to one for $t' \geq t_b$ and zero for $t' < t_b$. $t'$ is the time coordinate of a conscious being in the macroworld and $t_b$ is the time, in this coordinate system, when the final boundary condition occurs, initiating the last of the backward causal chains on the micro-system. Let us assign $t_a$ as the time of the earlier measurement in the EPR paradox. A description of the micro-system, valid at all times except $t_a \leq t' \leq t_b$ becomes:

$$\psi(x,y,z,t)*[1- u_s(t'- t_b)] + \psi_R(x,y,z,t)* u_s(t'- t_b) \quad t \leq t_b \qquad (12)$$

where $\psi$ is the standard quantum mechanical solution for the micro-system and $\psi_R$ represents the full reality of the inaccessible past. Identity of particles with the space-time coordinates has been suppressed. Since $\psi_R$ can never be observed directly except at $t = t_b$, we have to speculate what it is at time $t \leq t_b$, consistent with the observations at time $t_b$. For the EPR paradox, which we examined above, $\psi_R$ is a doubly reduced wavefunction in spin, but it is also reduced in the spatial and time coordinates. The detection of the electron at a localized space-time forces this second reduction. That this representation doesn't work at intermediate times of $t'$, between the two times of measurements in the EPR paradox, means that the wavefunction at these times is not Lorentz-invariant. Since change is taking place differently than the flow of time, the state does not have a well-defined existence at intermediate times. Only after all of the change is completed must the wavefunction be Lorentz-invariant. This is why the time of the last of the boundary conditions, $t_b$, appears and not the time of the earlier boundary condition. This well-known property of special relativity, that intermediate states do not need to be Lorentz-invariant, is another indication that something separate from the progression of time may represent change. I know of no other interpretation of special relativity, which gives a different ontological meaning to an intermediate state not being Lorentz invariant.

The difference between the progression of the causal chain in equation 11 and the time dependence in equation 12 is an essential feature of my model. Equation 11 describes a causal chain connected by an infinite series of events represented by a continuous path of particles moving through space-time. In fact, the causal chain brings to reality the events, which define the paths of the particles. Causally before the particles are detected there isn't any path along which they move. The causal chain can advance either forward or backward in time. In contrast, equation 12 gives a picture of reality from the perspective of an outside observer in a world, which advances only forward in time. At the instant the observer's event becomes a reality everything in the past, which is caused by that event, also becomes a reality. This is the so-called reduction of the wavefunction. Since the word instant refers to the time of the observer, rather than the time of the micro system, causality is not traveling faster than the speed of light.

**IX.     Comparison with the Copenhagen Interpretation**

Our model of backward causation is closely aligned with Bohr's Copenhagen interpretation. However it also accommodates some of the criticism, which Einstein and others have concerning Bohr's interpretation. Using the terminology of Willem M. de Muynck[15] Bohr's interpretation is based on the following interconnected ideas: *contextualistic realism, strong correspondence principle, complementarity, Copenhagen indeterminism, and probabilistic description of individual object.* Following is a brief description of these ideas contrasted with ideas favored by Einstein and compared to our model.

*Contextualistic realism* claims that reality of a property of an object comes solely from its interaction with a measuring instrument. In contrast, Einstein felt that there should be a theory, which can describe objective reality independent of measurement. Our model explains *contextualistic reality* as the effect of causal chains going both forward and backward in time and initiated by the measurement, which is the first cause in the chain. This idea comes from the basic principles in section II, which defines all reality in space-time as events, which are part of a causal chain. Realism must be contextualized in terms of both the initial preparation of the quantum state and the later measurement, since both actions initiate causal chains into the quantum system. Einstein wouldn't have any problem with contextualizing realism in terms of the preparation, and in our time-symmetric model, measurement is treated the same as preparation.

*The strong correspondence principle* claims that quantum phenomena corresponds to classical terms and can be unambiguously communicated only by classical terms. This idea is closely aligned to *contextualistic reality* in that reality can only be described in conjunction with the classical measuring device. However the Bohr model has an ambiguity here between ontology and epistemology. Whereas Bohr claimed the reality comes from the measurement, the classical description of the measurement is fundamentally flawed since classical concepts are partially inadequate to explain quantum phenomena. In our model of backward causation, classical concepts such as momentum or position of a particle become a reality as a result of measurement and/or preparation. In addition the classical concept of waves in QM is mostly associated with potentiality, which is a different kind of reality subject to change from a future measurement. Hence the correspondence principle is valid in that classical concepts of particles and waves are accurate when applicable and interpreted correctly, and is not inherently flawed.

*Complementarity* claims that incompatible observables cannot simultaneously have precise values because of the incompatibility of the measuring arrangement for each observable. This also includes particle –wave complementarity. For example, measuring arrangements are incompatible for observing a unique classical path of a particle versus observing an interference of two or more paths. In our model a particle, which has a unique classical position and/or momentum is constrained by the existence of causal chain(s). A particle, which acts more like a wave, with wavelets simultaneously traversing multiple paths, is less constrained by existing causal chains. The wave nature is a potentiality, which is open to the effects of causes, which have not yet occurred. The reason there is incompatibility of simultaneously observing these different phenomena is because the situations are different. In conjunction with *complementarity* is *Copenhagen indeterminism*, which claims that the value of a measured observable cannot be an attribute prior to measurement. In contrast Einstein treated indeterminism as epistemic. He felt that reality has to be precise. Our model has Bohr's concepts in a modified form. In the observer's reference frame, the measured observable is not a reality until after the measurement. However, because of backward causation the past is in a state of becoming. The measured attribute becomes a reality for the object at earlier times. *Complementarity* is also modified. In the situation where the preparation chooses a precise value of one observable and the measurement chooses a precise value of another incompatible observable, the object acquires precise values of both observables for the time between preparation and measurement. For example, in section II is a description of Stephen Hawking's quasi-Copenhagen interpretation of a free particle, which has two precisely measured space-time locations. Whereas Hawking would claim the particle doesn't have a well-defined momentum in-between the two measurements, my model claims the particle's path is a unique straight world line connecting the two space-time points. The world line defines precisely both the magnitude and direction of the particle's momentum. Here, momentum and position are incompatible observables, but both have precise values for all times in-between the two measurements.

The Copenhagen interpretation interprets quantum mechanics as giving a *probabilistic description of individual objects* rather than a statistical description of an ensemble of identically prepared objects. Specifically the probability distribution is a reality for individual microscopic objects and not simply a lack of knowledge. Einstein would favor an epistemic statistical description. For him, the particular microscopic object has precise properties, even if they are not classical properties, and must be thought of quantum mechanically as one in a possible ensemble of identically prepared objects. Our model favors the Copenhagen interpretation on this point. The probabilistic description is ontological for a single particle until a measurement is made on it. The measurement modifies the probability via backward causation by giving the particle a more precise value. The probabilistic reality is a different kind of reality than the reality of a measurement. The probabilistic reality is not composed of events in space-time. Rather its existence comes from the initial boundary conditions and the conservation laws. The conservation laws require certain properties of nature to exist and be real even before events, associated with these properties, come into existence through causal chains. The conservation laws constrain the causal chains, but do not create the causal chains.

The Copenhagen interpretation has some undesirable features. For example, it postulates that the observer obeys different physical laws that the non-observer, which has been criticized as a form of vitalism, that life is different from matter. Our model retains differences between observer and the quantum system, but defines better what these differences are, and it isn't vitalism. The pertinent differences between observer and observed are stated in principles 5,6,and 7 in the section II. The closest idea to vitalism is that humans, as agents, are free to impose their causal chains on objects of their choosing. The Copenhagen interpretation claims that the act of observing a system changes it in a random fashion, instantaneously over an extended region (non-local). Instantaneous is a bad word since there is no unique definition of simultaneity for spatially separated events. This paper has solved this non-locality problem using backward causation. Specifically the changes, which take place in a measurement, satisfy the locality condition of special relativity, in which causal chains cannot propagate faster than the speed of light.

## X. Final Comments

This paper has addressed only a few of the outstanding interpretational problems of QM. Our model is capable of solving many others. For example, it can naturally understand all of the experiments involving delayed choice or quantum eraser[16], which most other interpretations struggle to understand. The most outstanding issue, which is important for our model, is defining the boundary between the microworld, where both QM is valid and backward causation occurs, and the macroworld where QM doesn't apply and only forward causation occurs. The identification of this boundary cannot come solely from deductive logic and initially must be left for experimenters to probe. The boundary is likely defined by processes in the surrounding environment rather than the inherent properties of composite objects, which exist at the boundaries.

There are many so-called models of backward causation[17,18], but to my knowledge none of them allow the past to be in a state of becoming. Because there is no sense of becoming there are no causal chains, which create the becoming. Furthermore, none of these models incorporates humans as agents, who can create a new causal chain in the microworld, principle #5. All of these models appear to take a block universe approach, in which everything (past, present, and future) already exists. Some of the authors of these models will object that they are not taking a block universe approach, but I find their arguments very weak.

Many physicists would like to think that QM pertains to all of nature, not just the microworld. Such a view, if carried to its natural conclusion, forces a more anti-realist view of nature. Mermin[19] in a revealing paper about QM points out that physics has no notion of the concept of *now* unlike consciousness, which can "position itself absolutely along a world line of the being that possesses it." This is true because consciousness is aware of some of the past, which is fixed and looks to the future, which is in a state of becoming. The boundary between the two is defined as *now*. Mermin's model of QM, the so-called "Ithaca interpretation of QM," which eliminates the concept of *now*, limits physical reality to correlations between one time and another. Mermin claims that the things that are correlated, correlata, don't have physical reality. For example, he is implying that an electron can never have a definite momentum. Because correlata are not physically real there is no *now* or boundary between correlata, which are fixed and those in a state of becoming. Using backward causation my model can eliminate the concept of *now* without rejecting the physical reality of correlata. There is no sharp boundary between the correlata at earlier times and later times because they are all in a state of becoming. In my model an electron, which doesn't have a definite momentum, may causally obtain from a future measurement a specific value of momentum through backward causation. Since causation is moving backward in time the electron may have a well-defined path of motion at the earlier time. Backward causation allows a more realistic interpretation of QM. In contrast, Mermin relegates the measurement process to a minor role having little or no effect on physical reality. Backward causation clarifies the significance the measurement process plays by its ability to bring past propensities to actualization.

One of the outstanding problems in physics is to resolve the apparent incompatibility of QM with general relativity, GR. Most people, who work on this issue, struggle with the meaning of time and causality. Carlo Rovelli[20] in his paper *Quantum Spaecetime: What do we know?* Claims that the following conceptual questions have no unambiguous consentual agreement on answers: What is matter? What is causality? What is the role of the observer in physics? What is time? What is the meaning of being somewhere? What is the meaning of now? What is the meaning of moving? Is motion to be defined with respect to object or with respect to space? Our model has given fairly definitive answers to some of these questions and may be the key to resolving the discrepancy between QM and GR. Unfortunately, almost everyone working on this problem assumes QM is universal. This is a probable mistake. Julian Barbour presents an example of this approach in his book *The End of Time*[21]. He views the whole Universe as being in a stationary state. Actually, he pictures many static universes each at a different instant of time. Taken together, they appear to give an appearance of change, but they are not causally linked. This leads Barbour to the belief that "time does not exist at all and that motion itself is a pure illusion". Lee Smolin[22] has evaluated Barbour's argument and finds it correct, although he disagrees with Barbour's conclusion. He argues that some of Barbour's initial assumptions are probably faulty.

My model of backward causation is a metaphysical model, speaking against such ideas as anti-realism and the illusion of time. The arguments in favor of my model are based on the principles of symmetry, simplicity, and consistency, all of which are highly valued. We have not made a mathematical formulation of this model. To do so will require a totally new parameter, the causal order parameter. No such parameter appears in standard physics equations and considerable thought should be given before embarking on such a program. This is primarily metaphysics, without any predictive power, and may not need a mathematical formulation. But it provides a conceptual framework for resolving many QM paradoxes and inconsistencies, and may be fruitful in making QM compatible with GR.

Acknowledgement: I initiated this work at the 1998 Faculty Summer Seminar in Christian Scholarship at Calvin College financed by The Pew Charitable Trusts. In particular I would like to thank John Polkinghorne for his encouragement and insight as the seminar speaker. I also wish to thank Robert B. Griffiths, Michael B. Weissman, and Kenneth B. Wharton for critique and suggestions.